# Optical Atomic Clock Comparison through Turbulent Air

Martha I. Bodine[1], Jean-Daniel Deschênes[2], Isaac H. Khader[1,3], William C. Swann[1], Holly Leopardi[1,3], Kyle Beloy[1], Tobias Bothwell[4], Samuel M. Brewer[1], Sarah L. Bromley[4], Jwo-Sy Chen[1], Scott A. Diddams[1,3], Robert J. Fasano[1], Tara M. Fortier[1], Youssef S. Hassan[1,4], David B. Hume[1], Dhruv Kedar[4], Colin J. Kennedy[1,4], Amanda Koepke[1], David R. Leibrandt[1], Andrew D. Ludlow[1], William F. McGrew[1], William R. Milner[4], Daniele Nicolodi[1,3], Eric Oelker[1,4], Thomas E. Parker[1], John M. Robinson[4], Stefania Romish[1], Stefan A. Schäffer[1], Jeffrey A. Sherman[1], Lindsay Sonderhouse[4], Jian Yao[1], Jun Ye[1,4], Xiaogang Zhang[1,3], Nathan R. Newbury[1] and Laura C. Sinclair[1]

[1]*National Institute of Standards and Technology, 325 Broadway, Boulder, CO 80305, USA*
[2]*Octosig Consulting, Québec, QC, G1V 0A6, Canada*
[3]*Department of Physics, University of Colorado Boulder, Boulder, CO 80309, USA*
[4]*JILA, University of Colorado, Boulder CO 803039, USA*

We use frequency comb-based optical two-way time-frequency transfer (O-TWTFT) to measure the optical frequency ratio of state-of-the-art ytterbium and strontium optical atomic clocks separated by a 1.5-km open-air link. Our free-space measurement is compared to a simultaneous measurement acquired via a noise-cancelled fiber link. Despite non-stationary, ps-level time-of-flight variations in the free-space link, ratio measurements obtained from the two links, averaged over 30.5 hours across six days, agree to $6\times10^{-19}$, showing that O-TWTFT can support free-space atomic clock comparisons below the $10^{-18}$ level.

Optical atomic clocks, with stabilities and accuracies now approaching $10^{-18}$ [1–10], have created new opportunities for precision measurements in physics. These include the redefinition of the SI second, relativistic geodesy, investigation of possible variations in fundamental constants, and searches for dark matter, among others [11–24]. These applications require comparisons between clocks, and they have motivated optical atomic clock comparisons both within the same laboratory [1–4,15–17,25–28] and over fiber-optic links between laboratories [15,16,19,20,29–31]. However, recent progress in the development of high-performance portable atomic clocks [20,32–34], as well as continued interest in links between airborne or spaceborne clocks [35–50], highlights the need for methods of comparing atomic clocks over free-space links. Ideally, such methods would have residual instabilities and inaccuracies below those of the clocks themselves, despite the inevitable presence of atmospheric turbulence and platform motion.

To this end, we have explored frequency-comb-based optical two-way time-frequency transfer (O-TWTFT) across km-scale distances through turbulent air [42–48]. In previous experiments, our two optical clocks consisted of frequency combs phase-locked to cavity-stabilized lasers, which served as optical reference oscillators. With these clocks, O-TWTFT was used to compare the optical phase of two 195-THz (1535-nm) oscillators to below 10 milliradians [42], to compare their relative frequencies to fractional instabilities below $10^{-18}$ [42,43], and to actively synchronize two clocks to within a femtosecond, despite turbulence-induced signal fades and Doppler shifts due to motion [43–47].

In this work, we demonstrate the advancement of O-TWTFT in several ways. First, we demonstrate the capability of O-TWTFT to compare optical atomic clocks with transition frequencies that differ by over 90 THz. Second, we operate the O-TWTFT system in a new mode, allowing direct measurement of the frequency ratio of the two clocks. Finally, we confirm that O-TWTFT does not contribute any additional noise or systematic bias – the overall ratio uncertainty is limited by the clocks themselves, despite ps-level time-of-flight fluctuations in the transmitted optical timing signals.

The frequency ratio measurements discussed in this Letter were performed during a measurement campaign conducted by the Boulder Area Clock and Optical Network (BACON) collaboration [51]. This campaign compared three state-of-the-art optical atomic clocks: a ytterbium (Yb) lattice clock [4], an aluminum ion ($Al^+$) clock [5] and a strontium (Sr) lattice clock [7]. Here, we describe measurements of the ratio of $^{171}Yb$ and $^{87}Sr$ transition frequencies, obtained over 6 days using O-TWTFT across a free-space optical link. We compare these measurements with frequency ratio measurements simultaneously obtained using a conventional noise-cancelled fiber link [52,53]. Additionally, we evaluate the residual instability of the measurement network through a loopback test. We find that the frequency ratios measured by the free-space and fiber links agree to an uncertainty of $6\times10^{-19}$, and the instability of the loopback test reaches $1.5\times10^{-18}$ at a 1000-s averaging time.

The complexity of the experimental setup (Fig. 1) reflects the fact that the $^{171}Yb$ and $^{87}Sr$ atomic transition frequencies cannot be compared directly. The atomic transitions are separated in frequency by almost 90 THz, and the clocks themselves by 1.5 km. The Yb clock is located at the National Institute of Standards and Technology (NIST), while the Sr clock is located at the University of Colorado Boulder (CU). To enable the comparison, two optical frequency synthesis chains create phase-coherent connections between the atomic transition frequencies and the frequency combs within two associated O-TWTFT transceivers.

Inside the Yb Clock Lab at NIST, frequency-doubled light from a 259.1-THz (1157-nm) cavity-stabilized laser (the Yb clock laser) is maintained on resonance with the $^{171}Yb$ $6s^2$ $^1S_0$ – $6s6p$ $^3P_0$ transition frequency at 518.3 THz (578 nm). An optical frequency synthesis chain then connects this Yb atomic resonance frequency to the comb tooth frequencies of Comb A, located within the NIST O-TWTFT transceiver [43,46]. Similarly, inside the Sr Clock Lab, the 429.2-THz (698-nm) output frequency of a cavity-stabilized laser (the Sr clock laser) is maintained on resonance with the $^{87}Sr$ $5s^2$ $^1S_0$ – $5s5p$ $^3P_0$ transition. A second optical frequency synthesis chain connects the Sr atomic resonance frequency to the comb tooth frequencies of Comb B, located within the CU O-TWTFT transceiver.

To achieve a line-of-sight link over the city of Boulder, the two O-TWTFT transceivers were placed in a rooftop laboratory at NIST and in an 11[th] floor conference room at CU. Two low-insertion loss, free-space optical terminals [54] enabled bidirectional transmission of comb light between the two transceivers. Before transmission, the comb light was spectrally filtered to a 1.5-THz (12-nm) bandwidth centered at 192 THz (1560 nm). Additionally, to avoid receiver saturation, comb launch powers were attenuated by 3-10 dB from an initial in-band power of 5 mW.

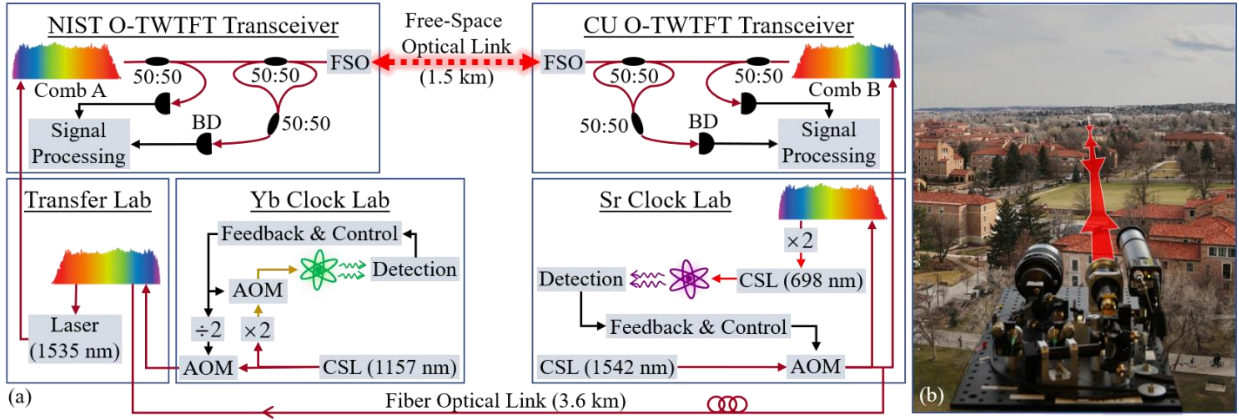

Figure 1: (a) In the Yb Clock Lab, frequency-doubled light from a 1157-nm cavity stabilized laser (CSL) is steered into resonance with the [171]Yb atomic transition via frequency shifts from an acousto-optic modulator (AOM). This steered 1157-nm light is sent to the Transfer Lab, where a fiber comb [55] transfers the frequency stability and accuracy of this 1157-nm light to a 1535-nm laser. This laser light is then sent to a rooftop laboratory where it serves as the reference for Comb A within the O-TWTFT transceiver. In the Sr Clock Lab, the frequency of a 698-nm CSL is steered into resonance with the [87]Sr atomic transition. The 698-nm light is steered by adjusting the frequency of a low noise 1542-nm CSL [56]. A frequency comb then acts to transfer the frequency adjustments of the 1542-nm light to the 698-nm light. The steered 1542-nm light also serves as the reference for Comb B within the O-TWTFT transceiver. On both sides of the free-space link, the heterodyne beat between local and remote comb pulses is detected and processed within the O-TWTFT transceivers. Additionally, the atomic clock frequencies are compared via a noise-cancelled fiber link. BD: balanced detector. (b) Photo from O-TWTFT transceiver at CU, looking towards NIST. A free-space optical (FSO) terminal with active beam steering maintains the bidirectional link between sites (photo edited to remove window frame).

The mathematical description of the optical frequency syntheses provides the following, exact relationship between the atomic frequency ratio and the repetition frequency ratio of combs A and B:

$$\frac{\nu_{Yb}}{\nu_{Sr}} = C_1 + C_2 \frac{f_{r,A}}{f_{r,B}}, \quad (1)$$

Here, $f_{r,A}$ and $f_{r,B}$ are the repetition frequencies of combs A and B, respectively; $\nu_{Yb}$ is the [171]Yb transition frequency; $\nu_{Sr}$ is the [87]Sr transition frequency; and constants $C_1$ and $C_2$ are known quantities—linear combinations of the optical and rf frequencies in the optical frequency synthesis chains. (See Supplemental Material [57] for a derivation.) We use O-TWTFT to measure the repetition frequency ratio $f_{r,A}/f_{r,B}$ in the optical domain, with high precision, despite time-of-flight variations from turbulence or building sway. Then, using (1), we compute the atomic frequency ratio from the measured repetition frequency ratio.

To avoid the inaccuracy associated with direct photodetection of pulse arrival times, O-TWTFT uses a linear optical sampling scheme to extract the relative frequencies of the combs [58]. The repetition frequencies of combs A and B, both near 200 MHz, are deliberately offset by $\Delta f_r = f_{r,A} - f_{r,B} \approx 2.3$ kHz. In each O-TWTFT transceiver, the pulse train of the local comb is mixed with the incoming pulse train from the remote comb. The resulting optical interference pattern is a series of interferograms, repeating at ~2.3 kHz. In both O-TWTFT transceivers, interferograms are digitized by an analog-to-digital converter (ADC) and processed to extract the exact times at which interferogram envelope maxima occurred. We refer to these extracted times as timestamps, denoting them $k_A$ and $k_B$ in the NIST and CU transceivers, respectively.

For precise frequency ratio measurements, avoiding the introduction of an external, cesium-based timescale is critical. Consequently, in both transceivers, the ADCs are clocked off the local comb's repetition frequency. One increment of ADC clock time $k$ is equal to the spacing between local comb pulses, and timestamps have units of local ADC clock-cycles, rather than seconds. Following [41,43,46], we label successive interferogram maxima with integer indices $p$ and write the timestamps as

$$k_A[p] = \frac{f_{r,A}}{\Delta f_r} p - \frac{f_{r,A} f_{r,B}}{\Delta f_r} T_{link}[p] \qquad (2)$$

and

$$k_B[p] = \frac{f_{r,B}}{\Delta f_r} p + \frac{f_{r,A} f_{r,B}}{\Delta f_r} T_{link}[p] , \qquad (3)$$

where $T_{link}$ is the varying time-of-flight across the reciprocal free-space path. Note that interferogram maxima are fit with sub-cycle accuracy, localized to 1/1024$^{th}$ of an ADC clock-cycle (roughly 5 ps). Consequently, timestamps $k_A$ and $k_B$ are non-integer. Although the timestamps have units of ADC clock-cycles, considering their approximate value in seconds can be helpful. If, for example, one divides each side of (2) by the local ADC clock rate, $f_{r,A}$, then the timestamps are scaled to nominal units of seconds, with an approximate spacing $\Delta f_r^{-1} \approx 435$ μs.

Were it not for ps-level, non-stationary variations in $T_{link}$, the combs' repetition frequency ratio could be obtained from a single set of timestamps, i.e. from either (2) or (3). Fortunately, the effects of time-of-flight variation may be eliminated by summing (2) and (3), giving

$$k_{sum}[p] = \frac{f_{r,A} + f_{r,B}}{f_{r,A} - f_{r,B}} p , \qquad (4)$$

as illustrated in Fig. 2. The slope of $k_{sum}$ versus $p$, obtained using ordinary least squares (OLS) fitting, is rearranged to find the comb repetition frequency ratio $f_{r,A}/f_{r,B}$ and then, using (1), the atomic frequency ratio. Equation (4) is very sensitive to changes in the repetition frequency. Consider a small fractional frequency change $(\delta f_{r,A}/f_{r,A})$ at the NIST site. With some manipulation, one finds the resulting fractional change in $k_{sum}$ is greatly magnified and is given by $(\delta k_{sum}/k_{sum}) \approx M(\delta f_{r,A}/f_{r,A})$, where $M = f_{r,A}/\Delta f_r \sim 10^5$. This magnification, which arises from the linear optical sampling technique [41,43,46], underlies the precision of O-TWTFT in determining the ratio.

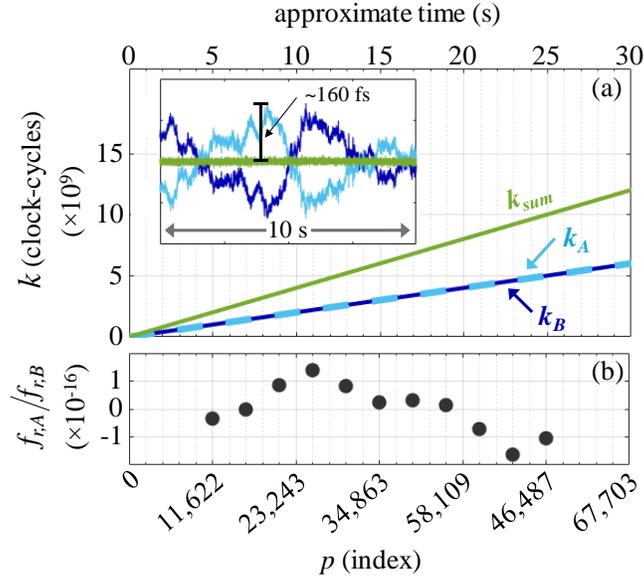

Figure 2: (a) Timestamps $k_A$ recorded at NIST and timestamps $k_B$ recorded at CU, as well as their sum $k_{sum}$. Inset: Expanded 10-s data segment with linear fits removed. Fluctuations in $k_A$ and $k_B$ are caused by atmospheric turbulence and building sway. When scaled from ADC clock-cycles to approximate time (scale bar), these can easily reach 100 fs in 1 s, but they are cancelled in $k_{sum}$. (b) Ratios of Comb A and Comb B repetition frequencies, offset from their mean. Each point is extracted from the slope fit to a single 10-s segment of $k_{sum}$.

O-TWTFT is inherently phase-continuous because it relies on a stable, uninterrupted local timescale (as defined by pulses of the local comb) to accurately record the timestamps. Here, the local time scales are extremely stable, since they are ultimately defined by atomic transitions. They are continuous as long as the optical frequency synthesis chains maintain the phase relationships between each clock laser and comb A or B. Note that phase continuity is not broken by signal fades due to turbulence or airborne debris. Signal fades along the free-space link can only reduce the total number of recorded timestamps; they cannot affect the stability or continuity of the local timescale [42]. However, both clock downtimes (i.e. times during which clock operation falls outside normal operating conditions) and phase slips in the optical frequency synthesis chains do interrupt the local timescale. Consequently, timestamps assigned before and after such an interruption have no relationship to one another, meaning that the slope of $k_{sum}$ in (4) may only be obtained from phase-continuous sections of data.

Identifying phase-continuous data involves two steps. First, clock downtimes, logged in the Sr and Yb clock labs, are flagged as discontinuities. Second, phase slips in the optical frequency syntheses are detected by scaling $k_{sum}$ to its approximate value in seconds and convolving it with a cycle-slip detection filter (see Supplemental Material [57]). Any sample-to-sample timing jumps greater than 1.9 fs are flagged, as 1.9-fs corresponds to a single $2\pi$ phase slip of the 578-nm Yb clock laser. Because the Yb clock laser has the highest frequency in the entire measurement system, this threshold is low enough to capture a phase slip anywhere within the optical frequency synthesis chains. The 30.5 hrs of data collected during the measurement campaign contained 1,201 segments of phase-continuous data, having mean, minimum and maximum durations of 98 s, 0.01 s and 46 min, respectively.

Within any segment of phase-continuous data, an unbiased, minimum-variance estimate of the slope of $k_{sum}$ depends on noise statistics. In the presence of white phase noise—i.e. uncorrelated noise between summed timestamps—this estimate is obtained by ordinary least squares (OLS) fitting. In the presence of white frequency noise—i.e. uncorrelated noise between successive values of the numerical derivative $k_{sum}[p] - k_{sum}[p-1]$—the mean of the derivative is the unbiased minimum-variance estimate. Additionally, because both noise types are zero-mean, these two estimators are unbiased for either noise type. In our data, white phase noise dominates at averaging times below 1 s (see Fig. S2) and white frequency noise (from the atom-steered clock lasers) at averaging times above 10 s. Between 1-s and 10-s averaging times, the noise is mixed. We use OLS to determine the slopes of 10-s sections within phase-continuous segments of data. (This requires rejecting segments which are phase-continuous for less than 10 s—less than 1% of the data.) These 10-s fits are overlapped by 80%, producing the frequency ratio results at 2-s intervals, as shown in Fig. 2b. The variance penalty incurred from using this estimation method, rather than the optimal method, is below 10%. (See Supplemental Material [57] for an expanded discussion.)

Between February and April of 2018, the atomic frequency ratio was measured on six days, producing 30.5 total hours of data. Individual ratio measurements are shown in Fig. 3, as fractional offsets from the ratio $R_{BIPM}$ calculated from the International Bureau of Weights and Measures (BIPM) values of 518 295 836 590 863.6 Hz and 429 228 004 229 873.0 Hz for Yb and Sr, respectively [59].

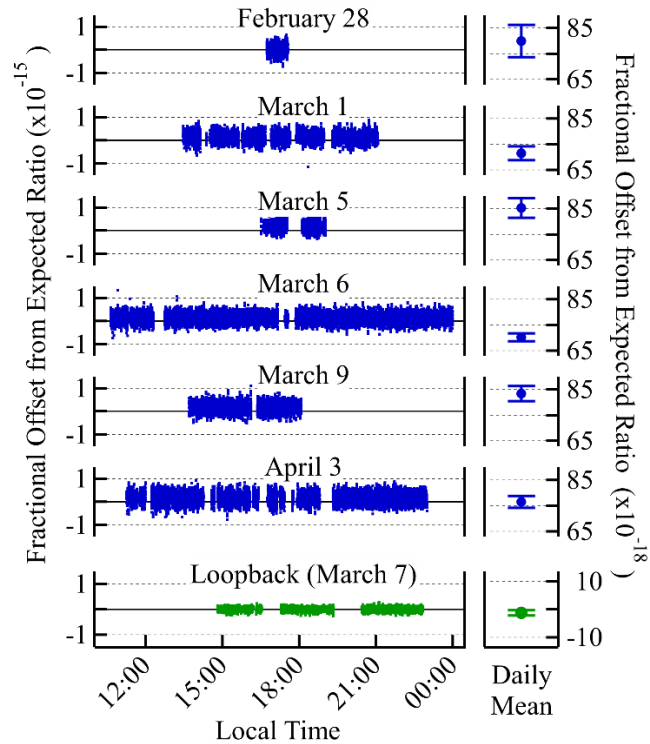

Figure 3: Measured atomic clock frequency ratios as a function of the local time in Boulder, CO. Ratios are plotted as fractional offsets: (*measured – expected*)/(*expected*). The fractional offset from the expected value for the atomic frequency ratio, $R_{BIPM}$, appears significantly displaced from zero but is well within the $\pm 6.4 \times 10^{-16}$ uncertainty of $R_{BIPM}$. The expected value of the loopback test is 1, since this test compares the frequency of the 1542-nm CSL to itself. Uncertainty bars show 1-sigma statistical uncertainty.

The overlapping Allan deviation for the ratio measurements from March 6 is shown in Fig. 4. The Allan deviation follows $3.8\times10^{-16}\tau^{-1/2}$, where $\tau$ is the averaging time in seconds, in accordance with the expected white frequency noise contribution from the clock lasers [51]. In other words, there is no evidence of additional noise due to the optical frequency synthesis chains or O-TWTFT. The statistical uncertainty of each daily mean in Fig. 3 is computed by extrapolating the measured Allan deviation to the full data set duration. (Systematic uncertainties are discussed in [51].)

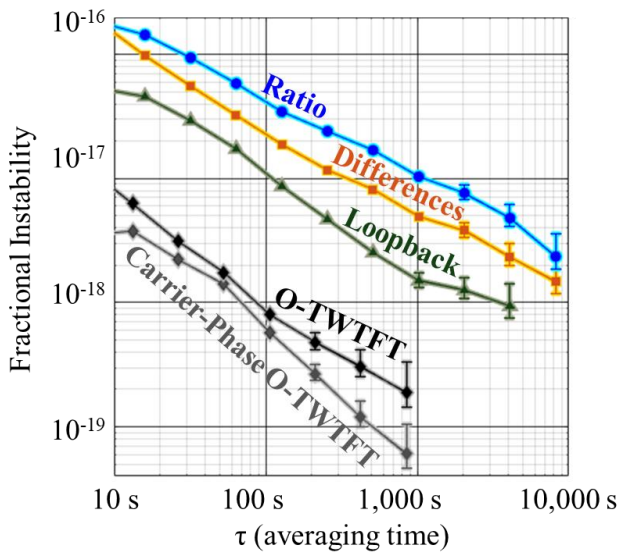

Figure 4: Fractional instabilities (overlapping Allan deviations) of atomic frequency ratio measurements from March 6, of the differences between O-TWTFT and fiber link measurements from March 6, and of the network loopback test. Also shown is the previously measured fractional instability (modified Allan deviation) of O-TWTFT itself, as well as that of carrier-phase O-TWTFT [42].

To determine an upper limit for any residual noise from the comparison network, we perform two additional analyses. First, we carried out a network loopback test, in which the origin of both optical frequency synthesis chains was the 1542-nm CSL in the Sr Clock Lab (Fig. 1a). The chain leading to the O-TWTFT transceiver at CU was unchanged, but the one at NIST was expanded to include the noise-cancelled fiber link between CU and NIST. This loopback test uses the measurement network to compare the frequency of the 1542-nm CSL to itself, thereby evaluating the cumulative statistical and systematic uncertainty of O-TWTFT, the phase-locked loops within the optical frequency synthesis chains, the noise-cancelled fiber links between laboratories, and the short, uncontrolled free-space or fiber paths within laboratories (other than those within the Yb Clock Lab). The mean of the 6-hr loopback test is offset from its expected value of 1 by $1.2\times10^{-18} \pm 0.9\times10^{-18}$, with the 1-sigma statistical uncertainty of $0.9\times10^{-18}$ taken from the last value of the Allan deviation (Fig. 4).

Second, we compare we compare $\nu_{Yb}/\nu_{Sr}$ ratio measurements obtained using O-TWTFT with simultaneous measurements obtained across a noise-cancelled fiber link [52,53]. We calculate the point-by-point differences between the O-TWTFT and fiber-link values at common measurement times, thereby removing common-mode Yb and Sr clock noise. This comparison assesses possible measurement errors between O-TWTFT and the noise-cancelled fiber link. Results, shown in Fig. 5, indicate no such errors, as the daily means of the differences are consistent with zero. Their weighted mean is -4.5×10$^{-19}$ ± 6×10$^{-19}$, well below the clocks' systematic uncertainties.

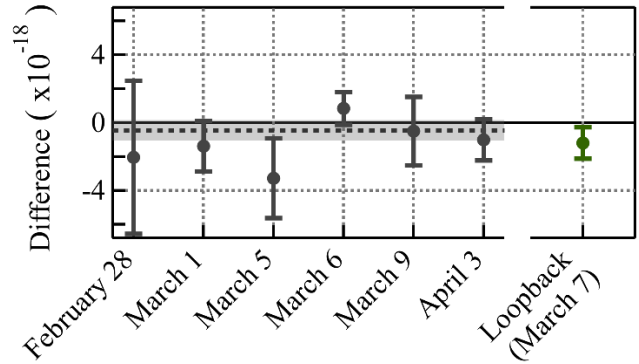

Figure 5. Differences between O-TWTFT and fiber measurements, along with their 1-sigma statistical uncertainties (error bars), their weighted average (dashed line) and its uncertainty (grey shading). Also shown is the offset of the network loopback test from its expected value of 1.

Also consistent with the removal of the common-mode clock noise, the instability of the differences is reduced compared to the instability of the ratio measurements themselves, although only by a factor of two (Fig. 4). This imperfect clock noise cancellation occurs because the two measurement approaches weight the raw data differently. The fiber link measurement uses a commercial lambda-type frequency counter, while the free-space analysis applies an overlapping parabolic filter to the raw frequency data. If the overlapping Allan deviation of the differences (Fig. 4) is extrapolated to the full 30.5-hr duration of the measurement campaign, the instability is 6×10$^{-19}$, consistent with the uncertainty quote above.

In conclusion, the frequency comb-based O-TWTFT measurements reported here agree to within 6×10$^{-19}$ with those recorded across a fiber noise-cancelled link, indicating that O-TWTFT can continue to support clock comparisons, even as clock accuracies cross below 1×10$^{-18}$. As transportable optical atomic clocks are developed [20,32–34], O-TWTFT will be able to support their comparisons for applications such as fundamental time metrology, relativistic geodesy measurements, or dark matter searches. With parallel work expanding the operation of O-TWTFT to longer links [45,60] and moving platforms [46,47], these comparisons could be made not only between mobile terrestrial clocks, but between future optical airborne and satellite-borne clocks as well.

Acknowledgements:

We thank Frank Quinlan and Ian Coddington for comments and Michael Cermak for technical assistance. We acknowledge funding from the DARPA DSO PULSE program and NIST.

# Optical Atomic Clock Comparison through Turbulent Air: Supplemental Material

**Supplemental 1: Allan deviations from all six measurement days**

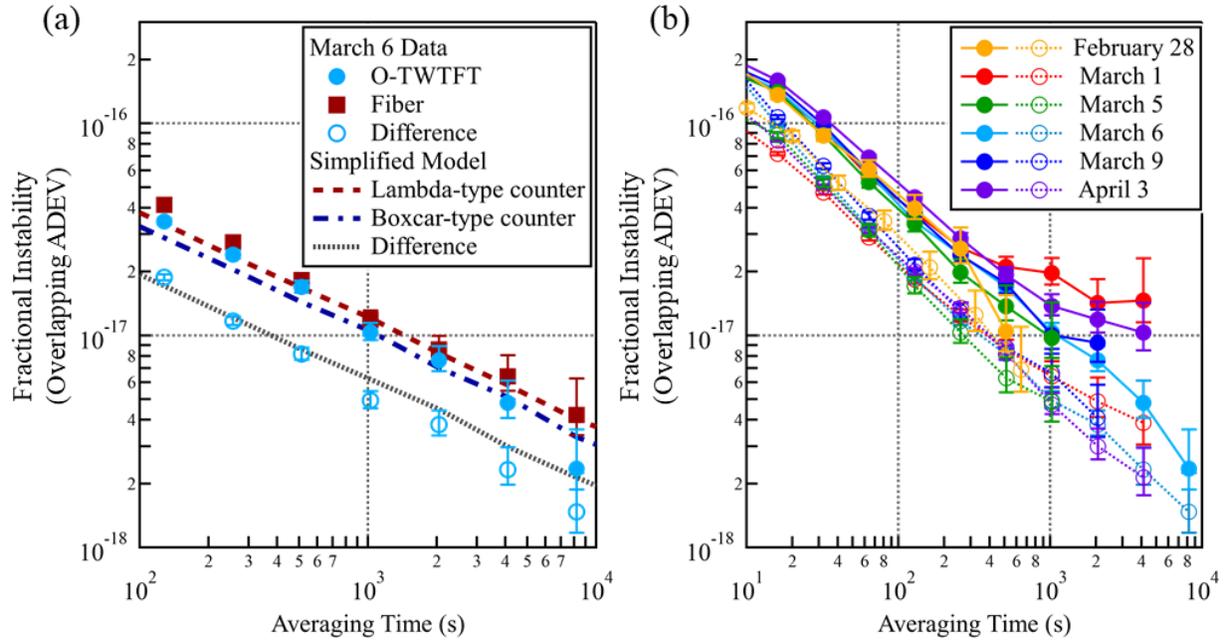

Figure S1: (a) Fractional instability (overlapping Allan Deviation) for March 6 for the free-space O-TWTFT measurement (solid blue circles), for the fiber-link measurements (solid red squares), and for their difference (open blue circles). As the two measurements are both dominated by the same white frequency noise from the clocks, one might expect the Allan deviations to fall on top of each other and the difference of the two measurements to have zero Allan deviation. This is not the case, since the two measurements use different weighting functions to estimate the frequency over the given averaging time. We plot the results of a simplified model, as dashed lines, that assumes identical white frequency noise for both measurements but differing weighting function, commonly referred to in terms of the counter type. (See Section S3 for more discussion of weighting functions.) In particular, note that because the fiber-link and free-space O-TWTFT measurements weight the frequency data differently, their individual ratio measurements differ, leading to an Allan deviation that is only lower by a factor of two from the ratio measurement itself. Data is shown only for averaging times at which the noise due to the probing of the atoms dominates, i.e. $\tau \geq 100$ s . (b) Fractional instability (overlapping Allan Deviation) for the frequency ratio measured via O-TWTFT (solid closed circles) and the difference between the O-TWTFT and fiber measurements (dashed open circles) for all six days of measurement. Note at long averaging times, there is larger scatter for the ratio measurement compared to the difference data, which is expected as any clock systematics are common mode for the difference data.

## Supplemental 2: Optical frequency synthesis chains

*S2.1 Ytterbium Optical Frequency Synthesis Chain*

The full ytterbium (Yb) optical frequency synthesis chain can be described mathematically as a relationship between the 518.3 THz, $^{171}$Yb neutral atom, $6s^2\ ^1S_0 - 6s6p\ ^3P_0$ unperturbed optical transition frequency [59] and the repetition frequency of Comb A in at NIST (Fig. 1). Following Fig. 1, the first algebraic link in this synthesis chain relates the Yb transition frequency $\nu_{Yb}$ to the optical frequency $\nu_{TL}$ of the 1157-nm light sent to the Transfer Lab, which is

$$\nu_{TL} = \frac{1}{2}(\nu_{Yb} - f_{corr,Yb}) + f_{FNC,1157}, \tag{5}$$

where $\nu_{Yb} - f_{corr,Yb}$ is the transition frequency during the experiment, and $f_{corr,Yb}$ represents postprocessing corrections applied to account for the difference between the transition frequency of the unperturbed Yb atom, and the transition frequency measured during the experiment in which the Yb atoms are subject to temperature fluctuations, lattice light shifts, etc. (Note, here and throughout the supplemental we have used the symbol $\nu$ to represent optical frequencies and $f$ to represent rf frequencies.) The offset $f_{FNC,1157}$ is half the demodulation frequency of the fiber noise-cancelled link connecting the Yb Clock Laboratory and the Transfer Laboratory. The factor of ½ is introduced because the 259.1 THz frequency from the cavity-stabilized laser in the Yb clock lab (Fig. 1) must be doubled to be on resonance with the $^{171}$Yb transition.

Again following Fig. 1, the second link in the Yb synthesis chain defines the optical frequency $\nu_A$ of the 1535-nm laser delivered to the NIST O-TWTFT transceiver, which is

$$\nu_A = \frac{N_{TL}}{M_{TL}}(\nu_{TL} - f_{CEO,TL} - f_{opt,1157}) + f_{CEO,TL} + f_{opt,1535} + f_{FNC,1535}. \tag{6}$$

Physically, the fiber frequency comb in the Transfer Laboratory spans the 64 THz gap between frequency $\nu_{TL}$ and frequency $\nu_A$. Mathematically, (6) may be derived from the well-known comb equation, with the addition of variable $f_{FNC,1535}$, which represents the frequency offset added by the fiber noise-cancelled link to the NIST O-TWTFT transceiver. Definitions and approximate values for all variables in equations (5), (6) and (3) are given in Table 1.

The third and final link in the Yb synthesis chain relates the repetition frequency $f_{r,A}$ of Comb A to the delivered optical frequency $\nu_A$ as

$$f_{r,A} = \frac{\nu_A}{N_A + r_{opt,A} + r_{CEO,A}} \tag{7}$$

where we again use the well-known comb equation and the integer $N_A$ is the mode number of the comb tooth that phase-locked to $\nu_A$. Comb A is fully self-referenced, meaning that the repetition frequency of the comb serves as the reference frequency for phase-locked loops fixing the comb's carrier-envelope offset (CEO) and optical offset frequencies. Consequently, its CEO frequency is written as $r_{CEO,A}f_{r,A}$ and the rf frequency of its optical phase lock to $v_A$ is $r_{opt,A} f_{r,A}$, where $r_{CEO,A}$ and $r_{opt,A}$ are user-selected rational fractions.

Combining all three links of the Yb optical frequency synthesis chain, i.e. Eq. (5) through (7), gives the following expression for the Yb transition frequency in terms of $f_{r,A}$,

$$v_{Yb} - f_{corr,Yb} = 2 f_{opt,1157} + 2 f_{CEO,TL} - 2 f_{FNC,1157}$$
$$- 2 \frac{M_{TL}}{N_{TL}} \left( f_{CEO,TL} + f_{opt,1535} + f_{FNC,1535} \right) + 2 \frac{M_{TL}}{N_{TL}} \left( N_A + r_{opt,A} + r_{CEO,A} \right) f_{r,A} . \tag{8}$$

All frequency terms on the right-hand side of (8), with the exception of $f_{r,A}$, are referenced to NIST hydrogen maser ST-15, and are known and fixed during the course of any experiment. The values in the correction term $f_{corr,Yb}$ are time-dependent and applied in post-processing.

Table I. Description of variables in Yb optical frequency synthesis chain. Comb TL is the frequency comb in the Transfer Lab. (Frequency comb mode numbers are exact integers.)

| Symbol | Value | Description |
| --- | --- | --- |
| $v_{Yb}$ | 518.3 THz (578.4 nm) | $^{171}$Yb neutral atom, 6s$^2$ $^1$S$_0$ – 6s6p $^3$P$_0$ unperturbed optical transition frequency, at the 811G104 geodetic marker at NIST |
| $f_{corr,Yb}$ | ~ Hz | Difference between true and measured $^{171}$Yb transition frequencies |
| $f_{FNC,1157}$ | 12.75 MHz | Frequency shift from the fiber noise-cancelled link between the Yb Clock Lab and the Transfer Lab |
| $v_{TL}$ | 259.1 THz (1156 nm) | Optical frequency delivered to the Transfer Lab |
| $f_{CEO,TL}$ | -80 MHz | CEO frequency of Comb TL in the Transfer Lab |
| $M_{TL}$ | 1,447,179 | Mode number of the Comb TL tooth phase-locked to optical frequency $v_{TL}$ |
| $f_{opt,1157}$ | -110 MHz | Offset frequency of phase-lock between comb tooth $M_{TL}$ and optical frequency $v_{TL}$ |
| $N_{TL}$ | 1,090,615 | Mode number of the Comb TL tooth to which the 1535-nm laser is phase-locked |
| $f_{opt,1535}$ | -20 MHz | Offset frequency of phase lock between comb tooth $N_{TL}$ and the optical frequency of the 1535-nm laser |
| $f_{FNC,1535}$ | 12.5 MHz | Frequency shift from fiber noise-cancelled link between the Transfer Lab and the NIST O-TWTFT transceiver |
| $v_A$ | 195.3 THz (1535 nm) | Optical frequency delivered to the NIST O-TWTFT transceiver |
| $f_{r,A}$ | 200.7358 MHz | Repetition frequency of Comb A in the NIST O-TWTFT transceiver |
| $r_{CEO,A}$ | -1/8 | CEO frequency (as a fraction of $f_{r,A}$) of Comb A |
| $N_A$ | 972,910 | Mode number of Comb A tooth phase-locked to optical frequency $v_A$ |

| $r_{opt,A}$ | -1/8 | Offset frequency (as a fraction of $f_{r,A}$) of phase lock between comb tooth $N_A$ and optical frequency $\nu_A$ |

*S2.2 Strontium Optical Frequency Synthesis Chain*

The strontium (Sr) optical frequency synthesis chain contains only two links, as shown in Fig. 1. The first defines the relationship between the 429.2 THz, $^{87}$Sr neutral atom, 5s$^2$ $^1$S$_0$ – 5s5p $^3$P$_0$ unperturbed optical transition $\nu_{Sr}$ and the optical frequency $\nu_B$ of the 1542-nm laser light delivered to Comb B at CU, which is

$$\nu_B = \frac{N_{Sr}}{M_{Sr}}\left(\nu_{Sr} - f_{corr,Sr} - f_{opt,698} - 2f_{CEO,Sr} + f_{FNC,698}\right) + f_{opt,1542} + f_{CEO,Sr} + f_{FNC,1542} \quad (9)$$

This equation is derived from the comb equation. The frequency $f_{FNC,698}$ is half the cumulative demodulation frequencies of the fiber noise-cancelled links connecting the Sr atoms to the frequency-doubled Comb Sr. The frequency $f_{FNC,1542}$ is half the demodulation frequency of the fiber noise-cancelled link connecting the 1542-nm CSL laser to the CU O-TWTFT transceiver. Note that the feedback loop inside the Sr Clock Laboratory uses the local fiber frequency comb to fix the phase relationship between the 1542-nm CSL and the 698-nm CSL which interrogates the Sr atomic lattice [7] (Fig. 1). Definitions and approximate values for all variables in (9) and (6) are given in Table 2.

The second link in the Sr synthesis chain defines the repetition frequency $f_{r,B}$ of Comb B in terms of the delivered optical frequency $\nu_B$. In analogy with Eq. (7) it is

$$f_{r,B} = \frac{\nu_B}{N_B + r_{CEO,B} + r_{opt,B}} \quad (10)$$

where $N_B$ is the mode number of the comb tooth number that phase-locked to $\nu_B$, and $r_{CEO,B}$ and $r_{opt,B}$ define the CEO and optical lock offset frequencies of the self-referenced Comb B, respectively. (See Table 2). The complete Sr optical frequency synthesis chain defines the Sr transition frequency $\nu_{Sr}$ as a function of the repetition frequency $f_{r,B}$ of Comb B as,

$$\nu_{Sr} - f_{corr,Sr} = f_{opt,698} + 2f_{CEO,Sr} - f_{FNC,698}$$
$$- \frac{M_{Sr}}{N_{Sr}}\left(f_{opt,1542} + f_{CEO,Sr} + f_{FNC,1542}\right) + \frac{M_{Sr}}{N_{Sr}}\left(N_{GT} + r_{opt,GT} + r_{CEO,GT}\right)f_{r,B} \quad (11)$$

Once again, all frequency terms on the right-hand side of (11), with the exception of $f_{r,B}$, are fixed relative to the ST-15 hydrogen maser during the course of the experiment. Time-dependent correction terms $f_{corr,Sr}$ are applied in post-processing.

Table II. Descriptions of variables in Sr optical frequency synthesis chain. Comb Sr is the frequency comb in the Sr Clock Lab. (Frequency comb mode numbers are exact integers.)

| Symbol | Value | Description |
| --- | --- | --- |
| $\nu_{Sr}$ | 429.2 THz (698.4 nm) | $^{87}$Sr neutral atom, 5s2 $^1S_0$ – 5s5p $^3P_0$ unperturbed optical transition frequency, at the JILAS1B60V1 geodetic marker at JILA. |
| $f_{corr,Sr}$ | ~ Hz | Difference between true and measured $^{87}$Sr transition frequencies |
| $f_{FNC,698}$ | 57 MHz | Cumulative frequency shift from fiber noise-cancelled links connecting the Sr atoms to the 698-nm CSL and connecting the 698-nm CSL to frequency-doubled Comb Sr |
| $f_{CEO,Sr}$ | 35 MHz | CEO frequency of Comb Sr in the Sr Clock Lab |
| $M_{Sr}$ | 1,716,882 | Mode number of the frequency-doubled Comb Sr tooth to which 698-nm CSL is phase-locked |
| $f_{opt,698}$ | 178 MHz | Offset frequency of phase lock between frequency-doubled comb tooth $M_{Sr}$ and the optical frequency of the 698-nm CSL |
| $N_{Sr}$ | 777,577 | Mode number of the Comb Sr tooth phase-locked to the 1542-nm CSL |
| $f_{opt,1542}$ | 35 MHz | Offset frequency of phase lock between comb tooth $N_{Sr}$ and the optical frequency of the 1542-nm CSL |
| $f_{FNC,1542}$ | 104.5 MHz | Frequency shift from the fiber noise-cancelled link between the Sr Clock Lab and the CU O-TWTFT transceiver |
| $\nu_B$ | 1542 nm | Optical frequency delivered to Comb B in the CU O-TWTFT transceiver |
| $f_{r,B}$ | 200.7335 MHz | Repetition frequency of Comb B in the CU O-TWTFT transceiver |
| $r_{CEO,B}$ | 1/8 | CEO frequency (as a fraction of $f_{r,B}$) of Comb B |
| $N_B$ | 968,437 | Mode number of the Comb B tooth phase-locked to optical frequency $\nu_B$ |
| $r_{opt,B}$ | 1/8 | Offset frequency (as a fraction of $f_{r,B}$) of phase lock between comb tooth $N_B$ and optical frequency $\nu_B$ |

*S2.3 Atomic Frequency Ratio*

The Yb and Sr optical frequency synthesis equations are combined to solve for the atomic transition frequency ratio $\nu_{Yb}/\nu_{Sr}$ as a function of the comb repetition frequency ratio $f_{r,A}/f_{r,B}$, yielding

$$\frac{\nu_{Yb}}{\nu_{Sr}} = C_1 + C_2 \frac{f_{r,A}}{f_{r,B}} \quad (12)$$

where

$$C_1 = \frac{2}{\nu_{Sr,BIPM}} \left[ \left( f_{CEO,TL} + f_{opt,1157} - f_{FNC,1157} + \frac{1}{2} f_{corr,Yb} \right) - \frac{M_{TL}}{N_{TL}} \left( f_{CEO,TL} + f_{opt,1535} + f_{FNC,1535} \right) \right] \quad (13)$$

and

$$C_2 = \frac{2}{\nu_{Sr,BIPM} M_{Sr}} \frac{M_{TL}}{N_{TL}} \frac{\left(N_A + r_{CEO,A} + r_{opt,A}\right)}{\left(N_B + r_{CEO,B} + r_{opt,B}\right)} \times \begin{bmatrix} M_{Sr}\left(f_{FNC,1542} + f_{opt,1542}\right) + \left(M_{Sr} - 2N_{Sr}\right)f_{CEO,Sr} \\ + N_{Sr}\left(f_{FNC,698} - f_{opt,698} - f_{corr,Sr} + \nu_{Sr,BIPM}\right) \end{bmatrix}. \quad (14)$$

Note that in order to solve expressions (8) and (11) for the atomic ratio in terms of the comb repetition frequency ratio, constants $C_1$ and $C_2$ must include either the International Bureau of Weights and Measures (BIPM) value of the Sr transition frequency $\nu_{Sr,BIPM}$ or the BIPM value of the Yb transition frequency $\nu_{Yb,BIPM}$ [59]. This is primarily due to the experimentally convenient use of the hydrogen maser to reference multiple, different rf offset frequencies within phase-locked loops in the optical frequency synthesis chains. Here we choose to use $\nu_{Sr,BIPM}$. The sensitivity of the ratio measurement to $\nu_{Sr,BIPM}$ is approximately $1.1 \times 10^{-18}$/kHz, while the inaccuracy of $\nu_{Sr,BIPM}$ is below 0.2 Hz. Therefore, any inaccuracy introduced to the ratio measurement by the inclusion of $\nu_{Sr,BIPM}$ is below $1 \times 10^{-20}$.

*S2.4 Maser inaccuracy*
All phase-locks and synthesized frequencies in both the Yb and Sr optical frequency synthesis chains were referenced to the same hydrogen maser. (The CEO and optical lock offset frequencies of combs A and B are exceptions, as these combs are fully self-referenced.) Hence, each frequency $f_x$ in (13) and (14) depends on the maser frequency, where the subscript $x$ stands for the different subscripts in Tables 1 and II. Given the values of the RF frequencies involved (see Table 1 and 2), the ratio of atomic frequencies has a sensitivity to the maser frequency of approximately $-46.5 \times 10^{-18}$/mHz. Therefore, for an uncertainty below $1 \times 10^{-18}$, the maser frequency offset must be known to below 21.7 µHz, corresponding to a fractional offset of $2 \times 10^{-12}$. The fractional offset $\varepsilon_{\text{ST-15 offset}}$ of maser ST-15 was measured during the experimental campaign [51] to well below $10^{-12}$ and was (coincidentally) $2 \times 10^{-12}$. Therefore, to accurately compute the atomic frequency ratio, each maser-referenced frequency $f_x$ which appears in in (13) and (14) was scaled in postprocessing as

$$f_x = f_{x,\text{nominal}}\left(1 - \varepsilon_{\text{ST-15 offset}}\right). \quad (15)$$

## Supplemental 3: Frequency ratio estimation from the O-TWTFT phase data

In order to estimate the frequency ratio, the slope of the summed timestamps, $k_{sum}[p]$, must be estimated. Without noise, the slope is obtained simply from the numerical derivative of $k_{sum}[p]$. However, in the presence of noise, there is an optimum estimator that produces a minimum-variance and unbiased result. As is often the case, and in the present system, both white phase noise and white frequency noise are present. At short timescales, additive white phase noise from O-TWTFT dominates (illustrated in Fig. S2), while at longer timescales, white frequency noise from the atomic clocks themselves dominates.

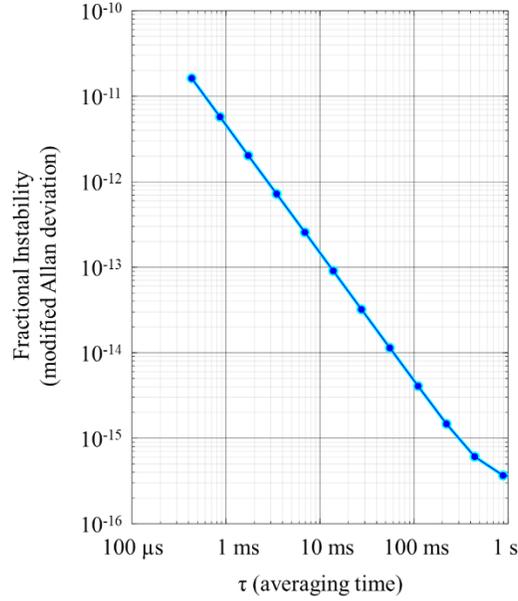

Figure S2: Fractional frequency instability (modified Allan deviation) of a phase-continuous segment of the summed timestamps taken on March 6. The instability follows $\tau^{-3/2}$, where $\tau$ is the averaging time in seconds, indicating that white phase noise is the dominant noise type at averaging times below 1 s.

It is well-known in the field of time and frequency that in the presence of white frequency noise, the optimal frequency estimator is simply a uniform average of each individual frequency measurements, while in the presence of white phase noise, the optimal estimator is a parabolic weighting of the individual frequency samples [61]. Here we outline the general derivation of the minimum-variance estimator for a mixture of both white phase and white frequency noise, using a generalized least-squares formalism. We then give the resulting weighting applied to our sampled data to achieve close-to-optimal estimation of the slope of (4).

The estimation of the slope of $k_{sum}$ is exactly analogous to the estimation of an oscillator's frequency from successive measurements of its phase $\phi$. In fact, $k_{sum}$ is related to the relative phase evolution of the frequency comb pulse trains at NIST and CU. Therefore, and to emphasize the general applicability of this analysis to time-frequency measurements, we outline the derivation of the optimal estimator for the case of estimating an oscillator's frequency from a series of phase measurements. We then can substitute appropriate quantities to relate this result to estimation of the slope of $k_{sum}$.

We assume we have measured a vector of $n$ phase samples given by $x[p] = (2\pi)^{-1} \phi(p\tau)$ at a given sample period $\tau$ where $p$ runs from 0 to $n-1$. We compute their numerical derivative to yield a vector of frequency estimates, $f[p] = \tau^{-1}(x[p] - x[p-1]) = A_{diff} x$, where $A_{diff}$ is the matrix that represents the linear transformation from phase samples to frequency samples,

$$A_{diff} = \frac{1}{\tau} \begin{bmatrix} 1 & -1 & 0 \\ 0 & \ddots & \ddots \\ 0 & 0 & 1 \end{bmatrix}. \tag{16}$$

The system is recast in matrix form as $f = X\beta + \varepsilon$, where $f$ is the $n$-element column vector of observed frequencies, $X$ is the $n \times m$ model matrix that relates these observations to the $m$-element vector of model coefficients, $\beta$, and $\varepsilon$ is the measurement noise with the associated covariance matrix $\Omega$. The optimal estimator is computed via generalized least-squares as $\hat{\beta} = (X^T \Omega^{-1} X)^{-1} X^T \Omega^{-1} f$. In our system, $\beta$ is simply the oscillator frequency, (i.e. $m=1$) and the observation matrix X is simply a $n$-element column vector with all elements equal to 1. The optimal frequency estimate, $\hat{f}$, is then $\hat{f} = (X^T \Omega^{-1} X)^{-1} X^T \Omega^{-1} f = Wf = \sum_{i=0}^{i=n-1} W_i f_i$, where we define the weighting vector $W = (X^T \Omega^{-1} X)^{-1} X^T \Omega^{-1}$. The covariance matrix $\Omega = \Omega_{wfn} + \Omega_{wpn}$ is the sum of a white frequency noise contribution, $\Omega_{wfn}$, which has elements on the main diagonal only, and the covariance matrix for the differentiated white phase noise, $\Omega_{wpn}$, which has off-diagonal elements. It is

$$\Omega = \sigma_f^2 I_n + (2\pi)^{-2} \sigma_\phi^2 \cdot A_{diff} A_{diff}^T \tag{17}$$

where $\sigma_f^2$ is the variance of the white frequency noise and $\sigma_\phi^2$ is the variance of the white phase noise at the given sampling rate. For white frequency noise only, i.e. $\Omega_{wpn} = 0$, we obtain the expected result of a boxcar uniform weighting vector, $W_i = n^{-1}$. For white phase noise only, i.e. $\Omega_{wfn} = 0$, we obtain a parabolic weighting that is given in Ref. [61]. For mixed noise, the optimal weighting vector (or function), W, lies between these two shapes.

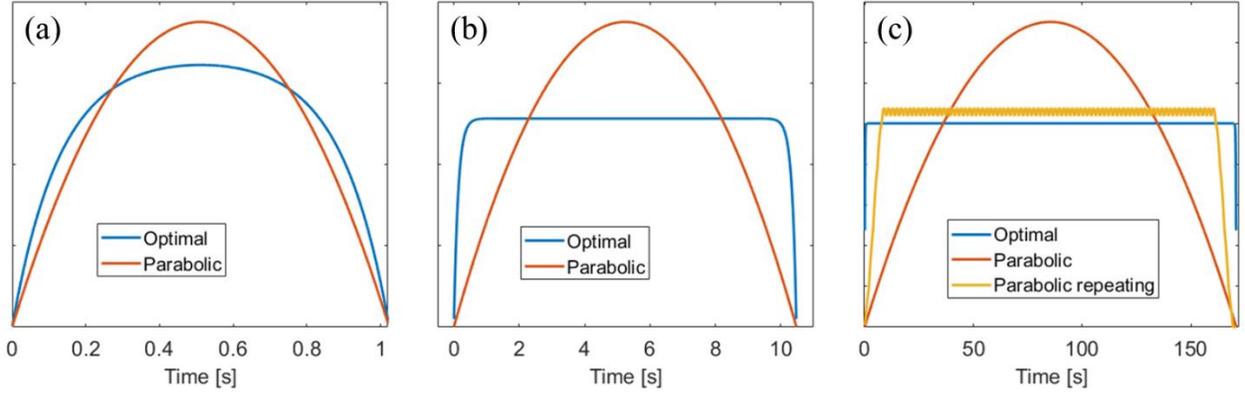

Figure S3: Weighting functions for a white phase noise corresponding to 2.5 fs timing jitter and white frequency noise corresponding to $5\times10^{-16}\tau^{-1/2}$ Allan deviation, at a sampling frequency of 2.3 kHz. (a) For a 1-second data segment, the minimum-variance, unbiased estimate is obtained from the optimal weighting (blue) compared to the parabolic weighting (red) for purely white phase noise. (b) For a 10-sec data segment, white frequency noise begins to dominate so that the optimal weighting shifts to a uniform boxcar. (c) For a 170-sec data segment, the optimal (blue) and parabolic (red) weighting differ dramatically, but the "repeating parabolic" weighting (yellow line) is close to optimal.

Figure S2 shows the different weighting functions for several dataset durations. It compares the optimal weighting function to a parabolic weighting function. Using values similar to those observed in our experiments, the sampling rate was $\tau =$ (2.3 kHz)$^{-1}$; the white timing jitter was 2.5 fs, and the white frequency noise corresponded to an Allan deviation of $4\times10^{-16}\tau^{-1/2}$. We can map this estimation over to Eq. (4) with the substitution of $x[p]\to k_{sum}[p]/2$, $\tau\to 1$, $(2\pi)^{-2}\sigma_\phi^2 \to \sigma_{ksum}=0.04$ and $\sigma_f^2 \to \sigma_{dksum}=0.00015$. (For example, the 2.5 fs timing jitter maps to a deviation in $k_{sum}/2$ of 0.04 by multiplying by the "effective timestep" of $f_r^2/(f_{r,A}-f_{r,B})$ associated with the linear optical sampling of one comb pulse train by the other).

From Fig. S3, it is clear that a purely parabolic weighting function is not optimal, particularly at longer dataset durations. However, the true optimum weighing function would have to be constantly recalculated since phase-continuous segments of the O-TWTFT data itself have random lengths due to the various phase slip or clock unavailability events. Instead, we take an approximate approach and implement a purely parabolic weighting (i.e. linear least squares fit of the slope) over 10-second intervals, which yields frequency estimates at 10-s timescales. We then overlap this calculation at 2-s intervals. The resulting vector of frequency estimates with 2-s spacing are finally averaged using a uniform boxcar weighing. The net result of these operations is a "repeating parabolic" weighting function that is close to a uniform weight but with a parabolic taper at the edge of each phase-continuous block, as shown in Fig. S1(c) for a 170-s dataset.

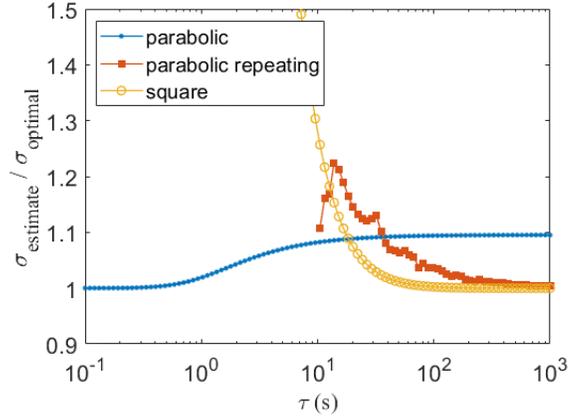

Figure S4: Standard deviation of the frequency estimate based on different weighting functions normalized to the optimal weighting function. For the analysis, only data durations longer than 10-s are analyzed, and the parabolic repeating weighting function is used, which applies a parabolic filter on 10-s data segments with 80% overlap.

Figure S3 quantifies the penalty for the several non-optimal weighting functions in terms of the increase in standard deviation as a function of the duration of phase-continuous data. A value of unity indicates no penalty, i.e. an optimal estimation given the noise. A value greater than unity indicates a penalty. As expected, the penalty for a uniform boxcar filter, appropriate for white frequency noise, is large for data set durations below 10 s. In contrast, the penalty for a parabolic filter, appropriate for white phase noise, grows for data set durations above 10 s, although only by a modest 10%. Nevertheless, the selected "repeating parabolic" weighting is generally a good compromise that covers the range of encountered data set durations.

We note that the calculation presented here assumes all O-TWTFT measurements are made with high signal-to-noise ratio (SNR), which minimizes the white phase noise. In reality, even over a short 1.5 km link, some data will be recorded at received powers close to the detection threshold. For these data, the white measurement noise can increase to greater than 10 fs [42], which would increase the penalty associated with a simple boxcar filter compared to that shown in Fig. S3. The use of the repeating parabolic filter avoids any such penalty.

**Supplemental 4: Phase-slip detection within the optical frequency synthesis chains**

Following a similar formalism to the one discussed above, phase-slip detection is implemented via a least-squares fit on $k_{sum}$ in a sliding window manner, over 2-s time windows. The 2-s time window is selected as a compromise between the level of noise reduction and temporal resolving power (i.e. how close in time two separate slip events can be detected). At 2 s, white phase noise dominates, and an ordinary least-squares fit is appropriate. We fit the data to a model that includes both the expected linear slope, along with a phase "step" in the center. (For completeness, we include a quadratic term to account for possible linear frequency drift during the 2-s window.) A non-zero estimate of the phase step would indicate a phase slip occurred at the center of the 2-s window.

This sliding-window least-squares fit is implemented by calculating the linear weights vector of the closed-form solution to the ordinary least-squares problem (OLS), and then convolving this weights vector with the observed $k_{sum}$. In other words, the fitting is formulated as a filtering operation, efficiently implemented using a fast Fourier transform algorithm.

Mathematically, we solve the equation $Y = A\beta$ where $Y$ is a column vector of the 2-s values of $k_{sum}$ scaled to phase measurements. The model matrix A is the usual Vandermonde matrix for a parabolic least-square fit, augmented with an extra row for the step discontinuity, and $\beta$ are the model parameters. The fourth element of $\beta$ is the fitted value for a phase slip. The model matrix is

$$A = \begin{bmatrix} 1 & t & t^2 & 2u\left(t - \frac{N}{2}\tau\right) - 1 \end{bmatrix} \tag{18}$$

or in alternative notation,

$$A = \begin{bmatrix} 1 & \tau & \tau^2 & -1 \\ 1 & 2\tau & (2\tau)^2 & -1 \\ \vdots & \vdots & \vdots & \vdots \\ 1 & (N-1)\tau & (N-1)^2\tau^2 & 1 \\ 1 & N\tau & (N\tau)^2 & 1 \end{bmatrix} \tag{19}$$

where $u(t)$ is the unit step function. The weights are then calculated in the usual manner for ordinary least squares via the Moore-Penrose pseudoinverse of matrix A to yield

$$W = \left(A^T A\right)^{-1} A^T . \tag{20}$$

The fourth row of the matrix W then corresponds to the weights for extracting the estimated magnitude of any discontinuity or phase slip. This weight vector is convolved with the data, and the result is compared to a threshold to identify a phase slip. These detected phase slips provide boundaries for the phase-continuous data blocks that are analyzed as discussed in the previous supplemental section.